\documentclass[fleqn,twoside]{article}
\usepackage{espcrc2}

\usepackage{graphicx}
\usepackage[figuresright]{rotating}

\newcommand{\AmS}{{\protect\the\textfont2
  A\kern-.1667em\lower.5ex\hbox{M}\kern-.125emS}}

\title{$\sigma_{DIS}(\nu N)$, NLO Perturbative QCD and
O(1 GeV) Mass Corrections}

\author{S. Kretzer\address[MCSD]{Physics Department and 
RIKEN-BNL Research Center, Bldg. 510a, Brookhaven 
National Laboratory, \\
Upton, New York 11973 -- 5000, U.S.A.}%
        \thanks{Funded in part by the U.S. Department of Energy contract DE-AC02-98CH10886
        and RIKEN.}and
        M. H. Reno\address{
        Department of Physics and Astronomy,
        University of Iowa, Iowa City, Iowa 52242 USA}\thanks{Funded in part by the U.S. Department         of Energy contract No. DE-FG02-91ER40664.} }
\begin{document}

\begin{abstract}
The deep-inelastic neutrino-nucleon cross section is one of the components
of few GeV neutrino interactions. We present here our results for neutrino-isoscalar
nucleon charged current scattering including 
perturbative next-to-leading order QCD corrections,
target mass corrections, charm mass and lepton mass corrections.
\vspace{1pc}
\end{abstract}

\maketitle

\section{INTRODUCTION}

Neutrino fluxes from cosmic ray interactions in
the atmosphere, and neutrino beams from long-baseline
neutrino oscillation experiments have in common that
a range of processes contribute to the cross section
and eventual event rates. In the few GeV region,
neutrino quasi-elastic scattering \cite{llsmith,strumia}, 
neutrino production
of one pion or a few pions \cite{rein,fogli,hmy,paschos}, and deep-inelastic
scattering (DIS)\cite{kr1,kr2} contribute to varying degrees \cite{lls}.
Ultimately, one would like to reduce the theoretical uncertainties
on the neutrino cross section in the few GeV region.

Discussed here are the contributions from charged current
(CC) deep-inelastic scattering. Similar results are
obtained for the neutral current case. We include lepton
masses, the target mass $M_N$ and charm mass corrections,
all in the context of the next-to-leading order (NLO)
QCD parton model \cite{kr1,kr2}. 

\section{FORMALISM}

The differential cross section is 
\begin{eqnarray} \nonumber
\frac{d^2\sigma^{\nu(\bar{\nu})}}{dx\ dy} &=& \frac{G_F^2 M_N
E_{\nu}}{\pi(1+Q^2/M_W^2)^2}\\ \nonumber
&\times & \left\{
(y^2 x + \frac{m_{\ell}^2 y}{2 E_{\nu} M_N})
F_1^{TMC} \right. \\ \nonumber
&+& \left[ (1-\frac{m_{\ell}^2}{4 E_{\nu}^2})
-(1+\frac{M_N x}{2 E_{\nu}}) y\right]
F_2^{TMC}
\\ \nonumber
&\pm& 
\left[x y (1-\frac{y}{2})-\frac{m_{\ell}^2 y}{4 E_{\nu} M_N}\right]
F_3^{TMC} \\  \nonumber
&+& 
\frac{m_{\ell}^2(m_{\ell}^2+Q^2)}{4 E_{\nu}^2 M_N^2 x} F_4^{TMC}
\\ \label{eq:nusig}
&-& \left. \frac{m_{\ell}^2}{E_{\nu} M_N} F_5^{TMC}
\right\}\ ,
\end{eqnarray} 
where target ($M_N$) and charged 
lepton masses ($m_\ell$) come into the limits of
integration via 
\begin{eqnarray}
 && \frac{m_\ell^2}{2M_N(E_\nu -m_\ell)}\leq   x\leq 1\\
&&a\ -\ b \leq  y\leq a\ +\ b \\ \nonumber
&&a  =  \Biggl[1-m_\ell^2\Biggl(\frac{1}{2M_NE_\nu x}+\frac{1}
{2E_\nu ^2}\Biggr)\Biggr]  \\ \nonumber &&\times (2+M_Nx/E_\nu
)^{-1}\\ \nonumber
&&b  =  \Biggl[\Biggl(1-\frac{m_\ell^2}{2M_NE_\nu
  x}\Biggr)^2-\frac{m_\ell^2}{E_\nu ^2}\Biggr]^{1/2}  \\ \nonumber
&& \times (2+M_Nx/E_\nu )^{-1}\ .\nonumber
\end{eqnarray}

\begin{figure}[htb]
\includegraphics[angle=270,width=15pc]{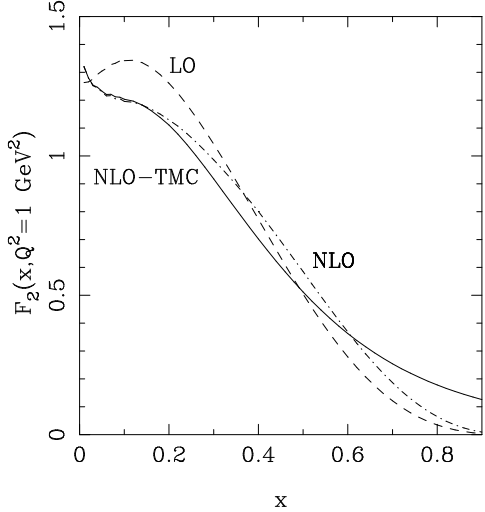}
\caption{The neutrino isoscalar nucleon structure function $F_2$ plotted
as a function of $x$ for $Q^2=1$ GeV$^2$. The solid line includes NLO QCD
corrections with target mass corrections. With $M_N\rightarrow 0$, the dot-dashed
line shows the NLO result
and the dashed line is the LO result.}
\label{fig:tmc}
\end{figure}

Target mass corrections (TMC) appear in the differential cross section
(Eq. (1)), in the limits of integration, and implicitly in 
$F_i$. The structure functions depend on the Nachtman variable \cite{nacht}
$\xi$ defined by
\begin{equation}
\xi = \frac{2 x}{1+\sqrt{1+\frac{4 M^2 x^2}{Q^2}}} \ .
\end{equation}
The Nachtman variable $\xi$ reduces to $x$ in the $M^2/Q^2\rightarrow 0$ limit.
In addition, one notes that the parton lightcone variable $p^+$
is a simple rescaling of the proton lightcone variable $P^+$ by
$p^+=\xi P^+$, however, $p^-\neq \xi P^-$ because $M_N\neq 0$.
The result is that the hadronic tensor describing the weak
interaction with the proton is not a simple rescaling of the
partonic tensor \cite{aot}. This leads to $M^2/Q^2$ corrections.
Further $M^2/Q^2$ corrections are attributed to the intrinsic
transverse momentum of the partons \cite{EFP}. Together, these can be derived
via the operator product expansion \cite{GP,DGP} and lead to, for example,
the target mass corrected structure function $F_2^{TMC}$ in terms of
the standard structure functions neglecting $M_N$:
\begin{eqnarray}
\label{eq:tmc}
F_2 ^{TMC} &=& 2\, \frac{x^2}{\rho^3}{{\cal F}_2(\xi ,Q^2)\over \xi}\\
\nonumber
&   +& 
12\,\frac{M_N^2}{Q^2}\,\frac{x^3}{\rho^4}
\int_\xi^1 d\xi '{ {\cal F}_2(\xi ',Q^2)\over \xi'}\\ \nonumber
&   + &
24\, \frac{M_N^4}{Q^4}\, \frac{x^4}{\rho^5} 
\int_\xi^1 d\xi ' \int_{\xi '}^1 { {\cal F}_2(\xi '',Q^2)\over
\xi ''}\ ,
\end{eqnarray}
where
\begin{eqnarray}
\rho^2& =& 1+\frac{4M^2x^2}{Q^2}\\
{\cal F}_2(\xi,Q^2)&=& q(\xi,Q^2)+\bar{q}(\xi,Q^2) + {\rm h.o.t.} 
\end{eqnarray}
Similar equations obtain for the other structure functions, and
are written in detail in Ref. \cite{kr2}.

\begin{figure}[htb]
\includegraphics[angle=270,width=15pc]{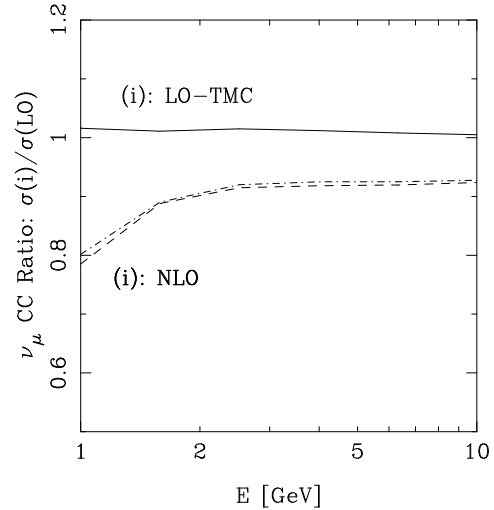}
\caption{The ratio of the LO with TMC (solid line),
NLO (dashed line) and NLO with TMC (dot-dashed line) to the LO cross section
with no TMC, as a function of incident neutrino energy for $\nu_\mu$-isoscalar
nucleon charged current scattering.}
\label{fig:murat}
\end{figure}

Charm quark mass corrections are incorporated via the `slow
rescaling variable' and charm mass corrected coefficient functions.
Our results reported in Ref. \cite{kr1} update those of Gottschalk
\cite{gotts}, and they are consistent with 
Ref. \cite{gkrcharm} in the $M^2/Q^2\rightarrow 0$ limit.

For the results here, we have added the additional constraint
that the final state hadronic invariant mass be larger
than a minimum value $W_{min}$ in order to avoid the 
double counting issues associated with combining the cross
sections for few pion processes and DIS. For results 
shown below, except as specified, 
\begin{equation}
W^2=Q^2\Bigl(\frac{1}{x}-1\Bigr)+M_N^2\geq W^2_{min}=(1.4\ 
{\rm GeV})^2\ .
\end{equation}
We remark that the issue of combining DIS and exclusive
processes is still an open question from a theoretical point
of view, although various practical schemes are employed by
a variety of Monte Carlo computer simulations.

We also note that we have included the full range of $Q^2$ in
our evaluation of the cross sections, where the strong coupling
constant and the parton distribution functions are frozen
below  a minimum value $Q^2=(1.3\ {\rm GeV})^2$. 
This extrapolation may or may not
be justified for the low energy cross sections, introducing an
additional theoretical uncertainty.

\section{RESULTS}

We first consider lepton mass corrections. For
$\nu_\mu N\rightarrow \mu X$, the LO DIS cross section
at 1 GeV is $\sim 60\%$ of the LO cross section
with $m_\mu=0$, denoted by $\sigma_0$. By 2 GeV, the CC cross section
is $\sim 95\%$ of $\sigma_0$. By contrast, the CC cross
section for $\nu_\tau N\rightarrow \tau X$ reaches 95\%
of $\sigma_0$ only at $E_\nu=1$ TeV.

\begin{figure}[htb]
\includegraphics[angle=270,width=15pc]{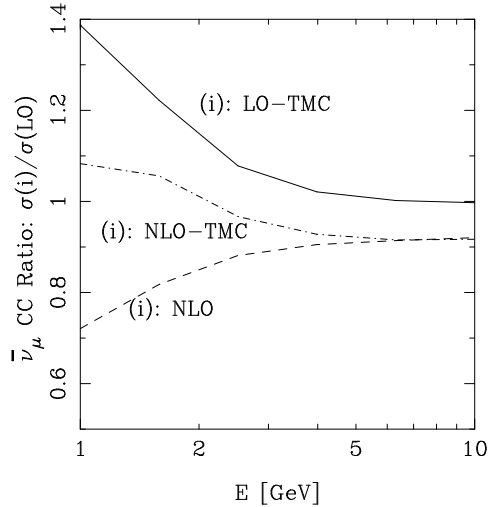}
\caption{As in Fig. 2, for $\bar{\nu}_\mu$-isoscalar nucleon 
charged current scattering.}
\label{fig:amurat}
\end{figure}

To discuss target mass corrections, we keep $M_M\neq 0$ fixed in the
expression
for the differential cross section, Eq. (1). By neglecting the TMC,
we set $\xi\rightarrow x$ and $F_i^{TMC}(\xi,Q^2)$ represented by
equations
like Eq. (\ref{eq:tmc}) are replaced by the uncorrected $F_i(x,Q^2)$.
The result for neutrino structure function $F_2^{TMC}$ for $Q^2=1$
GeV$^2$ is shown
in Fig. \ref{fig:tmc}. The curves for LO, NLO and the target mass
corrected NLO structure function are shown as a function of $x$, which
is implicit in $\xi(x)$. The NLO-TMC curve lies below the LO curve
at the smaller values of $x$, while the structure function with NLO
and
TMC is larger at large $x$. This comes from the fact that for
$x=1$ and $Q^2=1$ GeV$^2$, $\xi\simeq 0.65$. The parton distribution
functions are evaluated at $q(\xi, Q)$.

\begin{figure}[htb]
\includegraphics[angle=270,width=15pc]{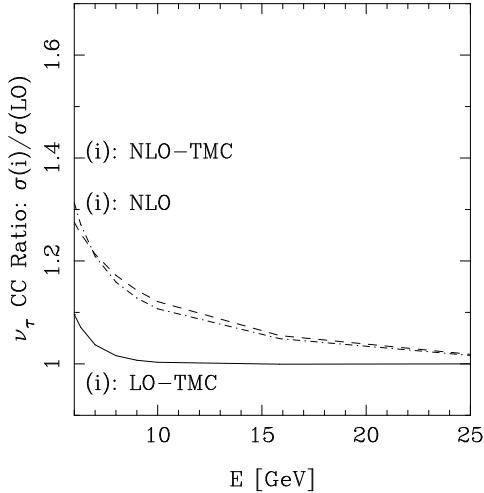}
\caption{As in Fig. 2, for $\nu_\tau$-isoscalar nucleon charged current
scattering.}
\label{fig:taurat}
\end{figure}

In 
Fig. \ref{fig:murat}, the solid line shows the ratio of the leading order
target mass corrected cross section to the leading order cross section
with $M_N=0$ as a function of incident muon neutrino energy.
The next-to-leading order (NLO) ratio and NLO with target mass corrections
ratio to the LO cross section is also shown in the figure. The target
mass corrections are not very important for $\nu_\mu N$ scattering.
A larger effect is seen in the case of $\bar{\nu}_\mu$
scattering
with nucleons
in Fig. \ref{fig:amurat}. The target mass corrections tend to moderate the NLO
corrections at low energies. 

\begin{figure}[htb]
\includegraphics[angle=270,width=15pc]{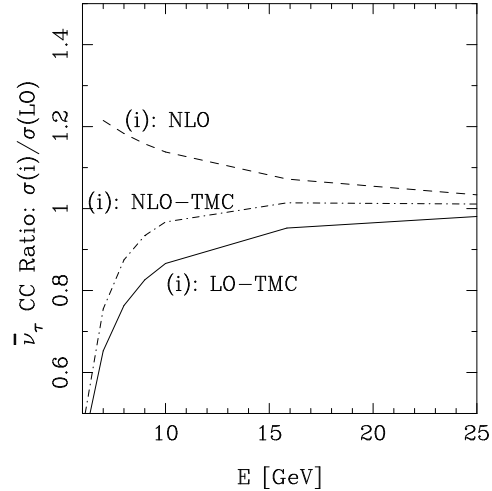}
\caption{As in Fig. 2, for $\bar{\nu}_\tau$-isoscalar nucleon charged current
scattering.}
\label{fig:ataurat}
\end{figure}

The different effects for neutrinos and
antineutrinos can be attributed to the fact that the cross sections
are dominated by the valence quarks. We note that
\begin{eqnarray}
\frac{d\sigma}{dy}(\nu N) &\sim & q(x,Q^2)\ \cdot \ 1 \\
\frac{d\sigma}{dy}(\bar{\nu} N) &\sim & q(x,Q^2)\ \cdot \ (1-y)^2 \ .
\end{eqnarray} 
The antineutrino cross section has a smaller average $y$ and therefore
a smaller $Q^2$ and larger $M^2/Q^2$ correction 
than the neutrino cross section.
Figs. \ref{fig:taurat} and \ref{fig:ataurat} show the same effect for $\nu_\tau$ and
$\bar{\nu}_\tau$ scattering with nucleons.

\begin{figure}[htb]
\includegraphics[angle=270,width=15pc]{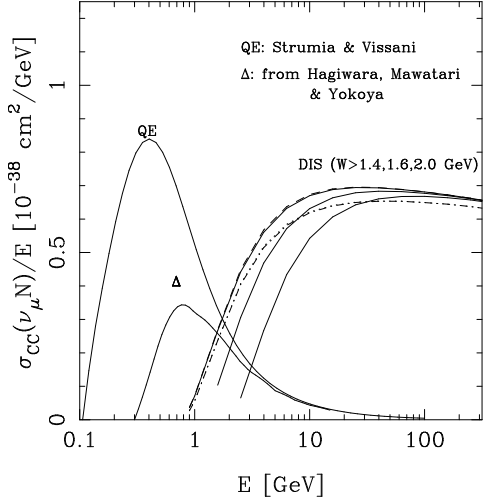}
\caption{Quasi-elastic (QE), resonant $\Delta$ and deep inelastic contributions
to $\nu_\mu N$ charged current scattering. The solid lines for the DIS contribution
are LO, with $W>1.4,\ 1.6$ and 2.0 GeV from left to right. The dot-dashed line is
the NLO corrected DIS cross section with $W>1.4$ GeV.}
\label{fig:sigmu}
\end{figure}

\begin{figure}[htb]
\includegraphics[angle=270,width=15pc]{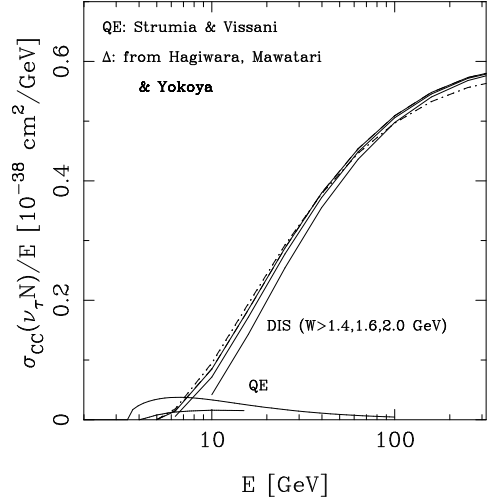}
\caption{As in Fig. 6, for $\nu_\tau N$ charged current scattering.}
\label{fig:sigtau}
\end{figure}

The context of the NLO and target mass
corrections can be seen from Figs. \ref{fig:sigmu}
and \ref{fig:sigtau}. In Fig. \ref{fig:sigmu}, we show
that DIS contribution together with the quasi-elastic cross section
and the cross section for $\Delta$ production. The quasi-elastic
(QE) neutrino-isoscalar nucleon cross section is evaluated using
the parameters summarized in Ref. \cite{strumia} by Strumia and
Vissani. The contribution coming from $\Delta$ production
has been evaluated by a number of authors. Here, we use the recent
results of Ref. \cite{hmy}, in which the cross section is
evaluated for $W<1.4$ GeV.  The left-most solid curve labeled
DIS is the LO evaluation of the DIS cross section for $W>1.4$ GeV.
The dashed line overlaying the solid line includes TMC. The dot-dashed
line shows the NLO corrected cross section with $W>1.4$ GeV. TMC
to the NLO curve are indistinguishable from the NLO curve.

In addition to the DIS curve for $W>1.4$ GeV, we also show for
reference
the DIS curves for $W>1.6$ GeV and $W>2.0$ GeV at leading order.
These two curves give a hint to the uncertainty in the 
procedure for combining inclusive (DIS) with exclusive
(QE and $\Delta$) cross sections. These other two DIS cross 
sections should be combined with their respective
$\Delta$ cross sections with $W<1.6,\ 2.0$ GeV to get the full
inelastic cross sections. 

The corresponding figure for $\nu_\tau$-isoscalar nucleon scattering
is shown in Fig. \ref{fig:sigtau}. The production of $\Delta$ is
much less significant in this case.

\section{DISCUSSION}

The range of DIS cross sections
values in Figs. \ref{fig:sigmu} and \ref{fig:sigtau}
as a function of the minimum value of $W$ gives a hint of the
range of uncertainty in the DIS contribution to the total
neutrino cross section. 
Target mass corrections are not important
on this scale of uncertainty, although they certainly are important
in limited regions of phase space. Target mass corrections
have a larger effect in antineutrino scattering
in the lower energy range of interest for $\bar{\nu}_\mu$, and for a larger
energy range for $\bar{\nu}_\tau$, partly compensating for the NLO enhancement
of the LO cross section.

Another approach to evaluating the few GeV DIS cross section 
by Bodek and Yang \cite{by} has been
to phenomenologically model electron-hadron electromagnetic
scattering in terms of rescaled parton distributions, then
to use the appropriate parton distribution combinations for
neutrino and antineutrino scattering with nucleons. 
A detailed comparison of this approach to the perturbative
QCD evaluation with the order GeV mass corrections would be useful,
in part to try to assess the importance of dynamical higher twist contributions
to the cross section. 

In principle, the transition from deep-inelastic to exclusive processes
in neutrino scattering can be modeled using duality arguments \cite{gb},
however, the details of a practical procedure are difficult to
determine \cite{lipari}. Our results here establish a reference standard
for our knowledge of the perturbative QCD prediction 
for neutrino-nucleon inelastic scattering including lepton mass,
charm mass and target mass corrections, an input to the theoretical
study of the
full picture of few GeV neutrino-nucleon interactions.

\end{document}